\documentclass[aps,prd,twocolumn,floatfix,showpacs]{revtex4-1} 

\usepackage{graphicx}

\usepackage{latexsym}

\begin{document}

\title{\boldmath Isospin breaking decay $\eta(1405)$\,$\to$\,$f_0(980)\pi^0$\,$\to$\,$3\pi$}
\author{N.~N.~Achasov\footnote{achasov@math.nsc.ru}\,$^1$,
A.~A.~Kozhevnikov\footnote{kozhev@math.nsc.ru}\,$^{1,2}$, and
G.~N.~Shestakov\footnote{shestako@math.nsc.ru}\,$^1$}
\affiliation{$^1$\,Laboratory of  Theoretical Physics,
S.~L.~Sobolev Institute for Mathematics, 630090, Novosibirsk,
Russia, \\$^2$\,Novosibirsk State University, 630090, Novosibirsk,
Russia}

\date{\today}

\begin{abstract} There are attempts in the literature to theoretically
explain the large breaking of isotopic invariance in the decay
$\eta(1405)$\,$\to$\,$f_0(980)\pi^0$\,$\to$\,$ 3\pi$ by the
mechanism containing the logarithmic (triangle) singularity, i.e.,
as being due to the transition $\eta(1405)\to(K^*\bar K+\bar
K^*K)\to(K^+K^-+K^0\bar K^0)\pi^0\to f_0(980)\pi^0\to3\pi$. The
corresponding calculations were fulfilled for a hypothetic case of
the stable $K^*$ meson. Here, we show that the account of the
finite width of the  $K^*$ ($\Gamma_{K^*\to K\pi}\approx50$ MeV)
smoothes the logarithmic singularities in the amplitude and
results in the suppression of the calculated decay width
$\eta(1405)\to f_0(980) \pi^0\to3\pi$ by the factor of  $6-8$ as
compared with the case of $\Gamma_{K^*\to K\pi}$\,=\,0. We also
analyze the difficulties related with the assumption of the
dominance of the $\eta(1405)\to(K^*\bar K+\bar K^*K)\to K\bar
K\pi$ decay mechanism and discuss the possible dynamics of the
decay $\eta(1405)\to\eta\pi \pi$.  The decisive improvement of the
experimental data on the $K\bar K$, $ K\pi$, $\eta\pi$, and
$\pi\pi$ mass spectra in the decay of the resonance structure
$\eta(1405/1475)$ to $K\bar K\pi$ and $\eta\pi\pi$, and on the
shape of the resonance peaks themselves in the $K\bar K\pi$ and
$\eta\pi\pi $ decay channels is necessary for the further
establishing the $\eta(1405)\to3 \pi$ decay mechanism.

\end{abstract}


\maketitle

\section{Introduction}

In seventies, a threshold phenomenon known as the mixing of
$a^0_0(980)$ and $f_0(980)$ resonances which breaks the isotopic
invariance, was theoretically discovered in Ref.~\cite{ADS79}, see
also Ref.~\cite{ADS81}. Recently, the interest in the
$a^0_0(980)-f_0(980) $ mixing has been renewed. New proposals for
searching it
\cite{Dz,AS97,KRS,KerT,CK,KudT,Gr,AK02,BHS,Ku,Bu1,Ko,Ha,
AT,Bu2,WYW,AS04a,AS04b,WZZ07,WZ08,Roca,Sek} have appeared, and the
results of the first experiments reporting its discovery with the
help of detectors VES \cite{Do08,Do11} and  BESIII \cite{Ab1,Ab2}
have been presented. The VES Collaboration was observed for the
first time the isospin breaking decay
$f_1(1285)$\,$\to$\,$\pi^+\pi^-\pi^0$ \cite{Do08,Do11}, the
proposal for searching it was put in Ref.~\cite{ADS79,ADS81}. The
BESIII Collaboration has obtained the indications on manifestation
of the $a^0_0(980)-f_0(980) $ mixing in the decays $J/\psi\to\phi
f_0(980)\to\phi a_0(980)\to\phi\eta\pi$ and $\chi_{c1}\to
a_0(980)\pi^0\to f_0(980)\pi^0\to\pi^+\pi^-\pi^0$ \cite{Ab1},
suggested for studies in Ref.~\cite{WZZ07,WZ08}. In another
experiment,  the  BESIII Collaboration  has measured the decays
$J/\psi\to\gamma\pi^+\pi^-\pi^0$ and
$J/\psi\to\gamma\pi^0\pi^0\pi^0$ and observed  the resonance
structure in the three pion mass spectra in the vicinity of 1.4
GeV with the width of about 50 MeV \cite{Ab2}. At the same time,
the corresponding $\pi^+\pi^-$ and $\pi^0\pi^0$ mass spectra in
the vicinity of 990 MeV (i.e. in the $K^+K^-$ and  $K^0\bar K^0$
threshold domain) possess the narrow structure with the width
about 10 MeV \cite{Ab2}. So, in this experiment, the isospin
breaking decay $J/\psi$\,$\to$\,$\gamma\eta(1405)$\,$\to$\,$\gamma
f_0(980)\pi^0$ followed by the transition
$f_0(980)\to$\,$\pi^+\pi^-(\pi^0\pi^0)$ was observed for the first
time \cite{Ab2} with the statistical significance exceeding
$10\sigma$. In the same experiment, the decay
$f_1(1285)/\eta(1295)$\,$\to$\,$\pi^+\pi^-\pi^0$ \cite{Ab2} was
also observed, with the  branching ratio by a factor of two lower
than that reported by VES \cite{Do11}.

The narrow resonancelike structure observed in the $\pi^+\pi^-$
and $\pi^0 \pi^0$ mass spectra in the decays
$\eta(1405)$\,$\to$\,$\pi^+ \pi^-\pi^0$, $\,\pi^0\pi^0\pi^0$ in
the $K^+K^-$ and  $K^0\bar K^0 $ threshold domain looks like the
structure expected to originate from the isospin breaking
$a^0_0(980)-f_0 (980)$ mixing \cite{ADS79}, i.e., due to the
transition $a^0_0(980)$\,$\to$\,$(K^+K^-+K^0\bar
K^0)$\,$\to$\,$f_0 (980)$\,$\to$\,$\pi\pi$ caused by the mass
difference of the  $K^+K^-$ and  $K^0\bar K^0$ intermediate
states. It should be recalled that the corresponding $S$ wave
amplitude responsible for the breaking of isotopic invariance, in
the region between $K\bar K$ thresholds (the width of this region
is about 8 MeV), turns out to be of the order of
$\sqrt{(m_{K^0}-m_{K^+})/m_{K^0}}$ \cite{ADS79,FN1}, but not
$(m_{K^0}-m_{K^+})/m_{K^0}$, i.e. by the order of magnitude
greater than it could be expected from the naive considerations.
It is natural to expect  the relative magnitude of the isospin
violation to be suppressed outside the $K\bar K$ threshold region,
i.e., at the level of  $(m_{K^0}-m_{K^+})/m_{K^0}$. To the first
approximation, one can neglect this and  the similar not really
calculable contributions.

The mechanism of the breaking of isotopic invariance in the decay
$\eta(1405)$\,$\to$\,$f_0(980)\pi^0$\,$\to$\,$3\pi$ is similar to
the mechanism of the $a^0_0(980)-f_0(980)$ mixing in that it is
caused by the transition $\eta(1405)$\,$\to$\,$(K^+K^-+K^0\bar
K^0)\pi^0$\,$\to$\,$f_0(980)\pi^0$\,$\to$\,$3\pi$. Its amplitude
does not vanish due to the nonvanishing mass difference of $K^+$
and $K^0$ mesons, and turns out to be appreciable in the narrow
region between the $K^+K^-$ and  $K^0\bar K^0 $ thresholds.

The aim of the present work is the elucidation of the possible
mechanism of the decay $\eta(1405)\to f_0(980)\pi^0\to\pi^+\pi^-
\pi^0$. There are attempts in the literature to theoretically
explain this decay as being due to the mechanism that includes the
logarithmic (triangle) singularities \cite{WLZZ12,ALQWZ12,WWZZ13},
i.e., due to the transition $\eta(1405)\to(K^*\bar K+\bar K^*K)\to
K\bar K\pi^0\to f_0(980)\pi^0\to\pi^+\pi^-\pi^0$. We pay attention
to the fact that in the cited works the vector $K^*(892)$ meson in
the intermediate state was considered to be stable, and show that
the account of the finite width of $K^*$,
$\Gamma_{K^*}\approx\Gamma_{K^*\to K\pi}\approx50$ MeV, smoothes
the logarithmic singularities in the amplitude resulting in the
suppression of the calculated width of the decay  $\eta(1405)\to
f_0(980) \pi^0\to3\pi$ by the factor of $6-8$ in comparison with
the case of $\Gamma_{K^*}$\,=\,0. We also analyze the difficulties
related to the assumption of the dominance of the decay
$\eta(1405)\to(K^*\bar K+\bar K^*K)\to K\bar K\pi$ and discuss the
possible dynamics of the decay $\eta(1405)\to\eta\pi\pi$. The
decisive improvement of the experimental data on the $K\bar K$, $
K\pi$, $\eta\pi$, and $\pi\pi$ mass spectra in the decays of the
resonance structure $\eta(1405/1475)$ \cite{PDG14} to $K\bar K\pi$
and $\eta\pi\pi$, and on the shape of the resonance peaks
themselves in the $K\bar K\pi$ and $\eta\pi\pi $ decay channels is
necessary for the further establishing the $\eta(1405)\to3 \pi$
decay mechanism.

\section{Experimental data}

According to BESIII \cite{Ab2}, the mass and width of the
$\eta(1405)$ peak in the $\pi^+\pi^-\pi^0$ channel are
$1409.0\pm1.7$ MeV and  $48.3\pm5.2$ MeV, respectively, while the
branching ratio is
\begin{eqnarray}\label{Eq1-1}
BR(J/\psi\to\gamma\eta (1405)\to\gamma
f_0(980)\pi^0\to\gamma\pi^+\pi^-\pi^0)\nonumber
\\ =(1.50 \pm0.11\pm0.11)\cdot10^{-5}\,.\qquad\quad\quad
\end{eqnarray}
Comparing the above with the result of Particle Data Group (PDG)
\cite{PDG14},
\begin{eqnarray}\label{Eq1-2}
BR(J/\psi\to\gamma\eta(1405/1475)\to\gamma K\bar K\pi)\nonumber
\\ =(2.8\pm0.6)\cdot10^{-3}\,,\qquad\qquad\quad
\end{eqnarray}one gets
\begin{eqnarray}\label{Eq1} \frac{BR(J/\psi\to\gamma\eta(1405)\to
\gamma f_0(980)\pi^0\to\gamma \pi^+\pi^-\pi^0)}{BR(J/\psi\to\gamma
\eta(1405/1475)\to\gamma K\bar K\pi)} && \nonumber \\
=(0.53\pm0.13 )\%\,.\qquad\qquad\qquad\ \ \, \end{eqnarray} The
magnitude of this ratio tells us about very large breaking of the
isotopic invariance in the decay
$\eta(1405)$\,$\to$\,$f_0(980)\pi^0$. Guided by naive
considerations, this ratio is expected to be at the level of
$[(m_{K^0}-m_{K^+})/m_{K^0}]^2\lesssim10^{-4}$. Notice that, in
Eq.~(\ref{Eq1}), the magnitude of the forbidden by isotopic
invariance decay $\eta(1405)\to f_0(980)\pi^0$ is compared to the
magnitude of the main allowed decay  $\eta(1405/1475)\to K\bar
K\pi$ \cite{PDG14,KW,MCU,AM}.

To illustrate the observed breaking of isotopic invariance, the
BESIII Collaboration \cite{Ab2} gives the ratio
\begin{eqnarray}\label{Eq1-3} \frac{BR(\eta
(1405)\to f_0(980)\pi^0\to\pi^+\pi^-\pi^0)}{BR(\eta
(1405)\to a^0_0(980)\pi^0\to\eta\pi^0\pi^0)} \nonumber \\
=(17.9\pm4.2 )\%\,.\qquad\qquad\qquad
\end{eqnarray}
However, it is large in comparison with Eq.~(\ref{Eq1}) due only
to the fact that the isospin-allowed transition
$\eta(1405/1475)\to a^0_0(980)\pi\to\eta\pi^0\pi^0$ is small.
Really, using the PDG branching ratio
$J/\psi\to\gamma\eta(1405/1475)\to\gamma\eta\pi^+ \pi^-)$
\cite{PDG10} and the largest PDG value of $\Gamma(\eta(1405) \to
a_0(980)\pi)/\Gamma(\eta(1405)\to\eta\pi\pi)$ \cite{Ams}, the
BESIII Collaboration \cite{Ab2} estimated $BR(J/\psi\to\gamma\eta
(1405)\to\gamma a^0_0(980)\pi^0\to\gamma\eta\pi^0\pi^0)=
(8.40\pm1.75)\cdot 10^{-5}$. So, the ratio Eq.~(\ref{Eq1-3}) is an
unreliable characteristic of the isospin violation.

In what follows we  also use the notation $\iota\equiv\eta(1405)$
for brevity. Since the decay $\iota$\,$\to$\,$f_0(980)\pi^0$ is
measured in the radiative decay of the $J/\psi$ meson, then, when
analyzing the situation, it is natural to base the treatment on
the information about the decays $J/\psi$\,$
\to$\,$\gamma\iota$\,$\to$\,$\gamma K\bar K\pi,\,\gamma\eta\pi\pi$
\cite{PDG14,KW,MCU,AM,Sc,Ed1,Au1,B1,Au2,B2,B3,Ed2,Bo,B4}. However,
this information is rather scarce. The matters are further
complicated by the fact that the data
\cite{Sc,Ed1,Au1,B1,Au2,B2,B3,Ed2,Bo,B4} refer to the decays
$J/\psi$\,$\to$\,$\gamma\eta(1405/1475)$\,$\to$\,$\gamma K\bar
K\pi,\,\gamma\eta\pi\pi$, in which the resonance structure
$\eta(1405/1475)$ \cite{PDG14,KW,MCU,AM} may correspond to some
mixture of the overlapping states $\eta(1405)$ and $\eta(1475)$
[it is called sometimes $\eta(1440)$ in current literature]. In
the meantime, there is no  single established opinion concerning
the reality of two pseudoscalars and the dynamics of the decays
 $\eta(1405/1475)$\,$\to$\,$K\bar K\pi$ and  $\eta\pi\pi$
\cite{WLZZ12,ALQWZ12,WWZZ13,PDG14,KW,MCU,AM}.

\section{The decay \boldmath $\iota\to(K^*\bar K+\bar K^*K)\to K\bar
K\pi^0\to$ $\to f_0(980)\pi^0\to\pi^+\pi^-\pi^0$}

\begin{figure}
\includegraphics[width=6.5cm]{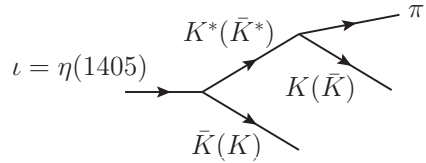}
\caption{\label{FigEtaKKpi}The diagram of the decay
$\iota\to(K^*\bar K+\bar K^*K)\to K\bar K\pi$.}\end{figure}

\begin{figure}
\includegraphics[width=7.5cm]{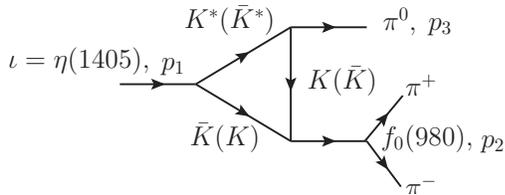}
\caption{\label{FigEta3pi}The diagram of the decay $\iota\to
f_0(980)\pi^0\to\pi^+\pi^-\pi^0$ via the $K^*\bar K+\bar K^*K$
intermediate states; $p_1$, $p_2$, $p_3$ stand for the 4-momenta
of particles participating in the reaction, $p_1^2=s_1$ being the
invariant mass squared of the $\iota$ resonance or of the final
$\pi^+\pi^-\pi^0$ system, $p_2^2=s_2=m^2_{\pi^+\pi^-}$ is the
invariant mass squared of the $f_0(980)$ or of the final
$\pi^+\pi^-$ system, $p_3^2=m^2_{\pi^0}$.}
\end{figure}

If the $\iota$ decays to  $(K^*\bar K+\bar K^*K)\to K\bar K\pi$ (see
Fig.~\ref{FigEtaKKpi}),
then, due to the final state interaction among $K$ and $\bar K$
mesons, i.e., due to the transitions $K^+K^-\to f_0(980)\to\pi^+
\pi^-$ and $K^0\bar K^0\to f_0(980)\to\pi^+ \pi^-$, the isospin
breaking decay $\iota\to(K^*\bar K+\bar K^*K)\to( K^+K^-+K^0\bar
K^0)\pi^0$\,$\to$\,$f_0(980)\pi^0$\,$\to$\,$\pi^+ \pi^-\pi^0$ is
induced (see Fig.~\ref{FigEta3pi}). It should be mentioned that
here we consider the effect of the isospin violation in the decay
$\iota\to\pi^+\pi^-\pi^0$ as being due solely to the mass
difference of the stable charged and neutral $K$ mesons.
The contributions from the production of the $K^+K^-$ and $K^0\bar
K^0$ pairs are not compensated completely. The smallest
compensation among them should naturally take place at the
invariant mass of the $\pi^+\pi^-$ system, $\sqrt{s_2}$, in the
region between the $K^+K^-$ and  $K^0\bar K^0$ thresholds.
\begin{figure}
\includegraphics[width=6.5cm]{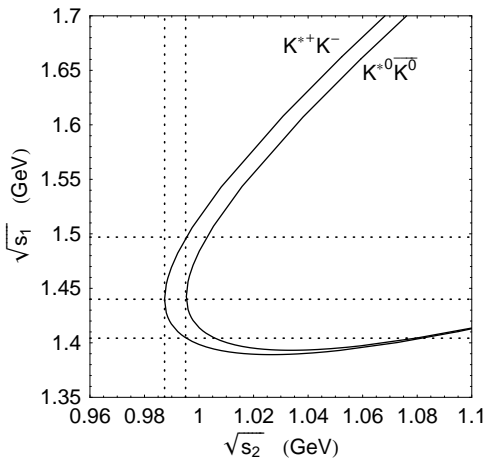}
\caption{\label{FigLogSing}Solid curves on the plane
$(\sqrt{s_2}\,,\sqrt{s_1} \,)$ show the location of the
logarithmic singularity of the imaginary part of the triangle
diagram shown in Fig.~\ref{FigEta3pi}, in the case of the
$K^{*+}K^-$ and  $K^{*0}\bar K^0$ intermediate states. The dashed
vertical lines show the $K^+K^-$  and  $K^0\bar K^0$ thresholds in
the variable $\sqrt{s_2}$ (i.e., its values equal to
$2m_{K^+}=0.987354$ and $2m_{K^0}=0.995344$ GeV). The dashed
horizontal lines correspond to the values of the variable
$\sqrt{s_1}$ equal to 1.404, 1.440, and 1.497 GeV. At
1.404\,GeV\,$<\sqrt{s_1}< $\,1.497\,GeV the logarithmic
singularity, in the case of the $K^{*+}K^-$ intermediate state, is
located at the values of $\sqrt{s_2}$ between the $K^+K^-$ and
$K^0\bar K^0$ thresholds, while in case of the $K^{*0}\bar K^0 $
intermediate state it does not go away from the $K^0\bar K^0$
threshold by farther than 6 MeV. At approximately
$\sqrt{s_1}$\,=\,1.440 GeV, the singularities reach the $K\bar K$
thresholds.}\end{figure}
However, there is some complexity in the present case. The fact is
that just in the region of the $\iota$ resonance all intermediate
particles in the loop of triangle diagram in Fig.~
\ref{FigEta3pi}, at the definite values of the kinematic variables
$\sqrt{s_1}$ and $\sqrt{s_2}$, can lie on their mass shells. This
means that in the hypothetic case of the stable $K^*$ meson the
logarithmic singularity appears in the imaginary part of the
triangle diagram \cite{AK1,AK2,AK3}. Figure \ref{FigLogSing} shows
the location of the logarithmic singularities for the
contributions of the $K^{*+}K^-$ and $K^{*0}\bar K^0$ intermediate
states. As is seen, in the $\iota$ resonance region, they are
located very close to the $K\bar K$ thresholds. For example, at
$\sqrt{s_1}$\,=\,1.420 GeV, the singularities from the $K^{*+}K^-$
and $K^{*0}\bar K^0$ intermediate state contributions in the
$\pi^+\pi^-$ mass spectrum take place at $\sqrt{s_2}$\,=\,0.989
GeV and  0.998 GeV, respectively (see Fig.~\ref{FigLogSing}).
Since the singularities located at different positions from the
charged and neutral intermediate states do not compensate each
other, the considered mechanism may seem to result in a
catastrophic violation of isotopic symmetry in the decay
$\iota\to\pi^+\pi^-\pi^0$. However, the accounting of the finite
width of the $K^*$ resonance, i.e., the averaging of the amplitude
over the resonance Breit-Wigner distribution in accord with the
spectral K\"{a}ll\'{e}n-Lehmann representation for the propagator
of the unstable  $K^*$ meson \cite{AK1,AK2,AK3}, smoothes the
logarithmic singularities of the amplitude and hence makes the
compensation of the contributions of the $K^{*+}K^-+ K^{*-}K^+$
and $K^{*0}\bar K^0+\bar K^{*0}K^0$ intermediate states more
strong \cite{FN2}.
\begin{figure}
\includegraphics[width=6.5cm]{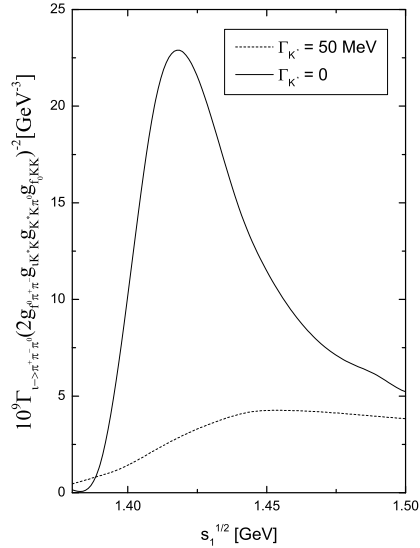}\caption{\label{G3pi-G3pi0}The illustration of
the influence of instability of the intermediate $K^*$ meson on
the calculated width of the decay $\iota\to(K^*\bar K+\bar
K^*K)\to K\bar K\pi^0\to f_0(980)\pi^0\to\pi^+\pi^-\pi^0$.}
\end{figure}
\begin{figure}
\includegraphics[width=9.5cm]{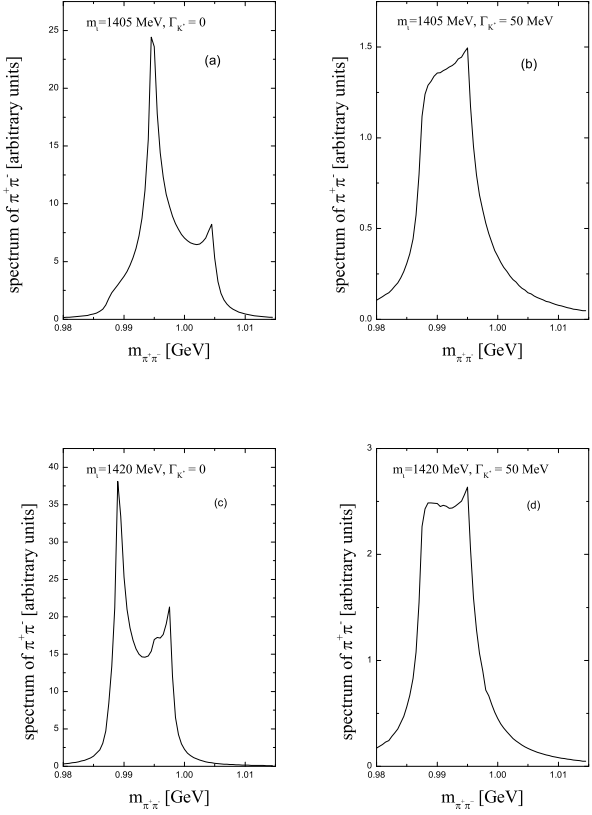}\caption{\label{FigSpec-pipi}The illustration of
the influence of instability of the intermediate $K^*$ meson on
the $\pi^+\pi^-$ mass spectra in the decay $\iota\to(K^*\bar
K+\bar K^*K)\to K\bar K\pi^0\to f_0(980)\pi^0\to\pi^+\pi^-\pi^0$.
The units are arbitrary but nevertheless the same for all
$\pi^+\pi^-$ mass spectra in (a) -- (d).} \label{FigSpec-pipi}
\end{figure}
This results in both the diminishing of the calculated width of
the decay $\iota\to\pi^+\pi^-\pi^0$ by a number of times in
comparison with the case of $\Gamma_{K^*\to K\pi}$\,=\,0, and  in
the concentration of the main effect of the isospin breaking in
the domain of the $\pi^+\pi^-$ invariant mass between the $K\bar
K$ thresholds. Figures \ref{G3pi-G3pi0},  \ref{FigSpec-pipi} show
the influence of allowing for the instability of  $K^*$ on the
energy dependent width $\Gamma_{\iota\to\pi^+\pi^-\pi^0}(s_1)$ and
on the mass spectra of the $\pi^+\pi^-$ system,
$d\Gamma_{\iota\to\pi^+\pi^-\pi^0}(s_1 ,s_2)/d\sqrt{s_2}$,
$\sqrt{s_2}=m_{\pi^+\pi^-}$. Figure \ref{G3pi-G3pi0} shows that in
the region 1.400 GeV $<\sqrt{s_1}<$ 1.425 GeV the calculated width
of the decay $\iota\to\pi^0\pi^+\pi^-$ is lowered by the factor of
$6-8$. The $\pi^+\pi^-$ mass spectra, see Fig.~\ref{FigSpec-pipi},
are distorted strongly. Notice that the nonzero experimental
resolution in the $\pi^+\pi^-$ mass (in the BESIII experiment
\cite{Ab2} -- it was about 2 MeV) would smooth the peaks in the
domain of singularity in Figs.~\ref{FigSpec-pipi} (a) and (c), but
the area under the curves would remain practically the same.

\begin{figure}
\includegraphics[width=10cm]{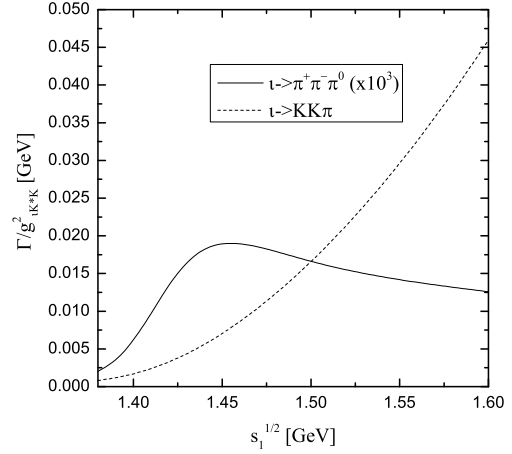}\caption{\label{FigRatioWidth}The dependence of the
$\iota\to(K^*\bar K+\bar K^*K)\to K\bar K\pi^0\to
f_0(980)\pi^0\to\pi^+\pi^-\pi^0$ and $\iota\to(K^*\bar K+\bar
K^*K)\to K \bar K\pi$ decay widths on the invariant mass of the
$\iota$ resonance $\sqrt{s_1}$ ($\Gamma_{K^*}=50$ MeV).}
\end{figure}

Shown in Fig.~\ref{FigRatioWidth} is the behavior of the
$\iota\to(K^*\bar K+\bar K^*K)\to K\bar K\pi^0\to f_0(980)\pi^0\to
\pi^+\pi^-\pi^0$ and  $\iota\to(K^*\bar K+\bar K^*K)\to K \bar K\pi$
decay widths against the invariant mass $\sqrt{s_1}$ of the $\iota$
resonance calculated at $\Gamma_{K^*}=50$ MeV. Both widths
demonstrate the strong dependence on $\sqrt{s_1}$. The ratio of
these widths is an important characteristic of the violation of the
isotopic invariance in the considered model. It does not depend on
the magnitude of the $\iota$ coupling with  $K^*\bar K$ ($g_{\iota
K^*\bar K}$), and its  order of magnitude is controlled by the
factor $[(m_{K^0}-m_{K^+})/m_{K^0}]\times(g^2_{f_0K^+K^-}/g^2_{f_0
\pi^+\pi^-})$ and decay kinematics. For the ratio of the widths in
Fig.~\ref{FigRatioWidth} averaged over the region 1.400
GeV\,$<\sqrt{s_1}<$\,1.425 GeV, one has
\begin{equation}\label{Eq2}
R=\frac{\overline{\Gamma}_{\iota\to\pi^+\pi^-\pi^0}}
{\overline{\Gamma}_{\iota\to K\bar
K\pi}}\equiv\frac{\langle\Gamma_{
\iota\to\pi^+\pi^-\pi^0}(s_1)\rangle} {\langle\Gamma_{ \iota\to
K\bar K\pi}(s_1)\rangle}\approx4\cdot10^{-3}\,.
\end{equation}
Now, using Eq.~(\ref{Eq1-2}) and (\ref{Eq2}) for evaluation of
$BR(J/\psi\to\gamma\iota\to\gamma f_0(980)\pi^0\to\gamma\pi^+
\pi^-\pi^0)$, one obtains
\begin{eqnarray}\label{Eq1-6}
BR(J/\psi\to\gamma\iota\to\gamma f_0(980)\pi^0\to\gamma\pi^+
\pi^-\pi^0)\nonumber \\ \approx R\times BR(J/\psi\to\gamma
\eta(1405/1475)\to\gamma K\bar K\pi)\nonumber \\
\approx1.12\cdot10^{-5},\ \qquad\qquad\qquad\qquad
\end{eqnarray}
in agreement with the data of  BESIII \cite{Ab2} given in
Eq.~(\ref{Eq1-1}).

The estimate Eqs.~(\ref{Eq1-6}) includes the assumption of
dominance of the $\eta(1405/1475)\to(K^*\bar K+\bar K^*K)\to K\bar
K\pi$ mechanism in the decay $\eta(1405/1475)\to K\bar K\pi$ to be
discussed below. Moreover, in view of the absence of the detailed
data, one forcedly assumes that  $\iota$ ($\eta(1405)$),
$\eta(1440)$, and the resonance complex $\eta(1405/1475)$
constitute the single object looking differently in various
channels.  Hence, the magnitude of  $BR(J/\psi\to\gamma\iota
\to\gamma f_0(980)\pi^0\to\gamma\pi^+ \pi^-\pi^0)$ given by
Eq.~(\ref{Eq1-6}) should be considered in the present model as the
upper estimate. See also remarks in Ref. \cite{FN3}.

\section{The decay \boldmath $\iota\to(K^*\bar K+\bar K^*K)\to K\bar K\pi$}

Guided by the data about the resonance complex $\eta(1405/1475)$
produced in the radiative decays of the $J/\psi$ meson one can
conclude that it decays to $K\bar K\pi$ with the probability of
about $80-90\%$ \cite{PDG14,KW,
MCU,AM,Sc,Ed1,Au1,B1,Au2,B2,B3,Ed2,Bo,B4}. The information of
about the contribution of the $K^*\bar K+\bar K^*K$,
$a_0(980)\pi$, $\kappa(800)\bar K+\bar\kappa (800)K$ intermediate
states to the decay $\eta(1405/1475)\to K\bar K\pi$ is contained
in two-particle mass spectra of the states $K\bar K$, $K\pi$, and
$\bar K\pi$. Available statistics of the $K\bar K\pi$ events are
not sufficient \cite{Sc,Ed1,Au1,B1,Au2,B2,B3}, so the quality of
the data does not permit one to reliably isolate the possible
contributions. To a very rough approximation it is assumed
\cite{PDG14,KW,MCU,AM} that the decay $\eta(1405/1475)\to K\bar
K\pi$ in the vicinity of 1475 MeV proceeds mainly via the $K^*\bar
K+\bar K^*K$ state. As for the region of 1405 MeV, it is
considered that it can proceed via the $a_0(980)\pi$ state
\cite{PDG14,KW,MCU,AM}, though the admixture of the $K^*\bar
K+\bar K^*K$ channel and even its dominance are discussed too
\cite{PDG14,KW,MCU,AM,Au2}. If, nevertheless, one admits dominance
of the $a_0(980)\pi$ channel, then it would be natural to expect a
rather sizeable signal from the decay $\iota\to
a_0(980)\pi\to\eta\pi\pi$ [$a_0(980)$ resonance is located near
the $K\bar K$ threshold and decays more intensively into $\eta\pi$
than into $K\bar K$]. In experiments, the decay
$J/\psi\to\gamma\iota \to\gamma\eta\pi\pi$ is seen
\cite{PDG14,KW,MCU,AM,Au1,Au2,Ed2,Bo, B4}, but it is small. See
the next section concerning this fact. One can definitely state
that the pointlike mechanism of the decay $\iota\to K\bar K\pi$
does not describe the data. So the assumption of the dominance of
the decay $\iota\to(K^*\bar K+\bar K^*K)\to K\bar K\pi$ cannot be
rejected as yet. The high statistics experimental studies of the
basic decay channels $\iota\to K\bar K\pi$ and
$\iota\to\eta\pi\pi$ are necessary for elucidation of the
situation.

\begin{figure}
\includegraphics[width=6cm]{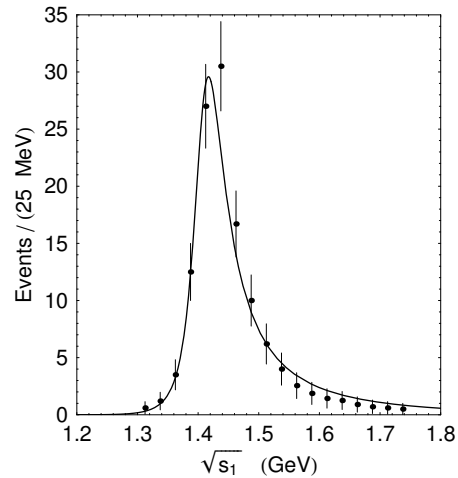}
\caption{\label{SpecKKpi}The $K\bar K\pi$ mass spectrum in the
decay $J/\psi\to\gamma\iota\to\gamma K\bar K\pi$ as the function
of the invariant mass of the $\iota$ resonance $\sqrt{s_1}$.
Points with error bars are the BES data \cite{B3}. The curve is
obtained in the $\iota\to(K^*\bar K+\bar K^*K)\to K\bar K\pi$
decay model. See the main text for more detail.}\end{figure}

In connection with the $\iota\to(K^*\bar K+\bar K^*K)\to K\bar
K\pi$ decay dominance we also want to pay attention to the
difficulty of using the simplest Breit-Wigner expressions for the
description of the $\iota$ resonance. For example, let us take the
recent BES data  \cite{B3} on the  $K\bar K\pi$ spectrum in the
decay $J/\psi\to\gamma\eta(1440)\to\gamma K\bar K\pi$, see
Fig.~\ref{SpecKKpi}, and fit them with the help of the standard
expression
\begin{eqnarray}\label{Eq1-7}
\frac{dN}{dm}=A(1-m^2/m^2_{J/\psi})^3BR(\iota\to K\bar K\pi;m)\,,
\end{eqnarray} where $m\equiv\sqrt{s_1}$, and
\begin{eqnarray}\label{Eq1-8}
BR(\iota\to K\bar K\pi;m)=\frac{2m}{\pi}\frac{m\Gamma_{\iota\to
K\bar K\pi}(m)}{|m^2_\iota-m^2-im\Gamma^{tot}_{\iota}(m)|^2}\,.
\end{eqnarray}
In the case of the total dominance of the $K^*\bar K+\bar K^*K$
channel, i.e., when
\begin{eqnarray}\label{Eq1-9}
\Gamma^{tot}_{\iota}(m)=\Gamma_{\iota\to K\bar
K\pi}(m)=\Gamma_{\iota\to(K^*\bar K+\bar K^*K)\to K\bar K\pi}(m),
\end{eqnarray}
the fit, shown in Fig.~\ref{SpecKKpi} with the solid line, gives
$\chi^2/n.d.f.=10/15$, $A=20$, $m_\iota=1.465$ GeV and  $g_{\iota
K^{*+}K^-}=6.91$ [hence $\Gamma_{\iota\to(K^*\bar K+\bar K^*K)\to
K\bar K\pi}(m_\iota)=448$ MeV, but the visible width of the peak
is essentially lower]. Our normalization is such that in the case
of the stable $K^*$ meson the coupling constant $g_{\iota
K^{*+}K^-}$ is related with the $\iota\to K^*\bar K+\bar K^*K$
decay width in accord with the expression
\begin{equation}\label{Eq6}
\Gamma_{\iota\to K^*\bar K+\bar K^*K}=\frac{g^2_{\iota
K^{*+}K^-}}{4\pi}\frac{8p^3_K}{m^2_{K^*}}\,,
\end{equation} where $p_K$ stands for the momentum of the $K$ meson in the
$\iota$ rest frame. If one evaluates the total $\iota\to K\bar
K\pi$ decay probability than instead of the expected value close
to 1 one would get
\begin{equation}\label{Eq1-10}
BR(\iota\to K\bar K\pi)=\int\limits^{3\mbox{\scriptsize{
GeV}}}_{1.3\mbox{\scriptsize{ GeV}}}BR(\iota\to K\bar
K\pi;m)dm\approx 0.34.\end{equation}The reason for this violation of
the normalization is the sharp $P$ wave growth of
$\Gamma_{\iota\to(K^*\bar K+\bar K^*K)\to K\bar K\pi}(m)$ with
increasing $m$ (see Fig.~\ref{FigRatioWidth}).

Recall that, in the case of the scalar mesons $\sigma(600)$,
$a_0(980)$, $f_0(980)$, their propagators obtained upon taking
into account the finite width corrections, satisfying the
K\"{a}ll\'{e}n-Lehmann representation and, due to this fact,
preserve the total decay probability normalization to unity
\cite{ADS80,AKi04}, see also Ref. \cite{AR14}. Unfortunately, we
have not yet succeeded in constructing the propagator for the
$\iota$ resonance, providing the desired normalization to unity,
as in the case of scalar mesons.

So, one can conclude that the fittings of the data on the $\iota$
resonance and the results of the determination of its parameters
from seemingly natural expressions should be considered as tentative
guesses.

\section{The decay  \boldmath $\iota\to a^0_0(980)\pi^0\to
f_0(980)\pi^0\to$ $\to\pi^+\pi^-\pi^0$}

The decay $\iota\to\pi^+\pi^-\pi^0$ can also proceed due to the
$a^0_0(980)-f_0(980)$ mixing \cite{ADS79}: $\iota\to
a^0_0(980)\pi^0$\,$\to$\,$f_0(980)\pi^0$\,$\to$\,$
\pi^+\pi^-\pi^0$. As a result, the $\pi^+\pi^-$ mass spectrum is
sharply enhanced in the region between the $K^+K^-$ and $K^0\bar
K^0 $ thresholds and looks very similar to the spectra shown in
Figs.~ \ref{FigSpec-pipi}(b) and \ref{FigSpec-pipi}(d). However,
it is difficult, with the help of this mechanism, to obtain the
magnitude of $BR(J/\psi\to\gamma \iota\to\gamma\pi^+\pi^-\pi^0)$
close to the experimental value Eq.~(\ref{Eq1-1}).

Let us take the data about the $a^0_0(980)-f_0(980)$ mixing
obtained by BESIII \cite{Ab1},
\begin{equation}\label{Eq1-14}
\xi_{af}=\frac{\Gamma_{a^0_0\to f_0\to\pi^+\pi^-}}{\Gamma_{a^0_0\to
\eta\pi^0}}=(0.31\pm0.16\pm0.143)\%.
\end{equation}
Notice that the upper limit on $\xi_{af}$ is $1.0\,\%$ at $90\,\%$
confidence level \cite{Ab1}. Let us also base the consideration on
the magnitude
\begin{eqnarray}\label{Eq1-14a}
BR(J/\psi\to\gamma\iota\to\gamma\eta\pi^+\pi^-)\qquad\quad \nonumber
\\ =BR(J/\psi\to\gamma\eta
(1405/1475)\to\gamma\eta\pi^+\pi^-)\nonumber \\
=(3.0\pm0.5)\cdot10^{-4}\qquad\qquad\ \ \ \
\end{eqnarray}  \cite{PDG14},
and let us consider the decay $\iota\to\eta \pi^+\pi^-$ as
proceeding via the $(a^+_0(980)\pi^-+a^-_0(980)\pi^+) $ intermediate
states. Then one obtains for $BR(J/\psi\to\gamma\iota
\to\gamma\pi^+\pi^- \pi^0)$:
\begin{eqnarray}\label{Eq1-15}
BR(J/\psi\to\gamma\iota\to\gamma\pi^+\pi^-\pi^0)\quad \nonumber \\
=\frac{\xi_{af}}{2}\,BR(J/
\psi\to\gamma\iota\to\gamma\eta\pi^+\pi^-)\nonumber\\
\approx(4.5\pm3.3)\cdot10^{-7}\,.\qquad\ \ \ \
\end{eqnarray} The central value in Eq.~(\ref{Eq1-15}) is by
approximately  30 times lower than the central value given by
Eq.~(\ref{Eq1-1}). However, the experimental uncertainties of the
data on $\xi_{af}$ are large, and one needs additional
measurements to make definite conclusions.

\begin{figure}
\includegraphics[width=6cm]{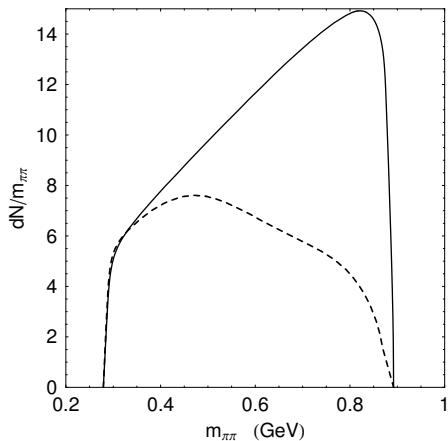}
\caption{\label{SpecEtapipi}The $\pi\pi$ mass spectrum in the
decay $\iota\to\eta\pi\pi$. The solid curve corresponds to the
$\iota\to a_0 (980)\pi \to\eta\pi\pi$ decay mechanism. The dashed
curve corresponds to the decay via the
$a^{\pm,0}_0(980)\pi^{\mp,0}$ intermediate states upon taking into
account the $S$ wave $\pi\pi $ final state
interaction.}\end{figure}

In general, the suppression of the decay
$J/\psi\to\gamma\iota\to\gamma \eta\pi^+\pi^-$ as compared with
the $J/\psi\to\gamma\iota\to \gamma K\bar K\pi$ one  \cite{PDG14}
is not directly related with the smallness of the  $\iota\to
a_0(908)\pi$ decay probability. Hence, the branching ratio
$BR(J/\psi\to\gamma \iota\to\gamma \pi^+\pi^-\pi^0)$, caused by
the $a^0_0(980)-f_0(980) $ mixing mechanism, can be few times
greater than that  given in Eq.~(\ref{Eq1-15}). The fact is that
the $a_0(980)\pi$ intermediate state in the  $\iota\to\eta\pi\pi$
decay channel can be hidden due to the destructive interference
with other contributions. As our estimates show, the interference
between $a_0(980)\pi$ and $\sigma(600)\eta$ intermediate states
can reduce the probability of the decay $\iota\to\eta\pi\pi$ by
the factor of about 1.5; see also Ref.~\cite{PaPi}. Besides, the
$S$ wave $\pi\pi$ final state interaction in the decay $\iota\to
a_0(980)\pi\to\eta\pi\pi$ is capable of suppressing its width by
the factor of approximately two. The possible influence of this
interaction on the $\pi\pi$ mass spectrum in the decay
$\iota\to\eta\pi\pi$ is shown in Fig.~\ref{SpecEtapipi}. So, the
estimate Eq.~(\ref{Eq1-15}) can be enhanced by the factor of
approximately three. If such a possibility is realized, it would
mean that the contribution of the $a^0_0(980)-f_0(980)$ mixing
mechanism can provide up to 30\% of the $\iota\to\pi^+\pi^-\pi^0$
decay amplitude.

The high statistics experimental investigations on both the form
of the mass spectrum of the  $\iota$ resonance in the $\eta\pi\pi$
decay channel and the $\eta\pi$ and  $\pi\pi$ subsystem mass
spectra in the region of $\iota$ peak could elucidate considerably
the production dynamics and the role of the $a_0(980)\pi $
intermediate state.

\section{Details of calculations}

To estimate the effect, we use the following expression for the
propagator of stable $K^*$ meson:
\begin{equation}\label{Eq4}
\frac{g_{\mu\nu}-k_\mu k_\nu/k^2}{m^2_{K^*}-k^2-i\varepsilon}\,.
\end{equation} It preserves the conservation of the unit spin in
the presence of interaction and the convergence of the triangle
diagram in Fig.~\ref{FigEta3pi} for the  intermediate states with
the specific charge. It should be stressed that the convergence or
divergence of the triangle diagram as well as of the $K\bar K$
loops in the case of the $a^0_0(980)\to(K^+K^-+K^0\bar K^0)\to
f_0(980)$ transition is not related with the effect under
discussion. The sum of the subtraction constants for the
contributions of the charged and neutral intermediate states in
the dispersion representation for the isospin breaking amplitude
should have the natural order of smallness  $\sim(m_{
K^0}-m_{K^+})$, and it cannot be responsible for the enhancement
of the symmetry violation in the vicinity of the $K^+K^-$ and
$K^0\bar K^0$ thresholds.

The contribution of the triangle diagram  in Fig.~\ref{FigEta3pi},
divided by the product of coupling constants $g_{\iota K^\ast
K}g_{K^\ast K\pi^0}g_{f_0KK}$ is
\begin{widetext}
\begin{equation}\label{tri1}
T\equiv
T(s_1,s_2,m^2,m^2_K)=\int\frac{d^4p_K}{(2\pi)^4}\cdot\frac{(p_1+p_{\bar
K})(p_\pi-p_K)-(s_1-p^2_{\bar K})(m^2_\pi-p^2_K)/p^2_{K^\ast}}
{(m^2-p^2_{K^\ast}-i\varepsilon)(m^2_K-p^2_K-i\varepsilon)(m^2_K-p^2_{\bar
K}-i\varepsilon)},
\end{equation}
\end{widetext}
where $s_1$, $s_2$, $m^2$ are, respectively, the invariant masses
squared of $\iota$, $f_0$, and $K^\ast$. The numerator of the
integrand contains polynomials $(m^2-p^2_{K^\ast})$,
$(m^2_K-p^2_K)$, and $(m^2_K-p^2_{\bar K})$ which cancel some poles
in the denominator. Hence the expression for $T$ reduces to the sum
of terms with three and two propagators each treated using the
Feynman parametrization:
\begin{eqnarray*}
\frac{1}{a_1a_2a_3}&=&2\int_0^1dx_1\int_0^{x_1}\frac{dx_2}{[a_1x_2+a_2(x_1-x_2)+a_3(1-x_1)]^3},\nonumber\\
\frac{1}{a_1a_2}&=&\int_0^1\frac{dx}{[a_1x+a_2(1-x)]^2}.
\end{eqnarray*}
After integration over $p_K$ the logarithmic divergences in the
two-propagator contributions cancel, and the resulting expression
can be represented in the following form:
\begin{widetext}
\begin{eqnarray}\label{T}
T&=&\frac{1}{16\pi^2}\left\{\left[s_1+m^2_\pi-m^2+2(m^2_K-s_2)+\frac{(s_1-m^2_K)(m^2_K-m^2_\pi)}{m^2}\right]C_3(s_1,s_2,m^2,m^2_K)-
\right.\nonumber\\&&\left.
\frac{(s_1-m^2_K)(m^2_K-m^2_\pi)}{m^2}\cdot
C_3(s_1,s_2,0,m^2_K)+\frac{s_1-m^2_K}{m^2}\left[C_2(s_1,m^2,m^2_K)-C_2(s_1,0,m^2_K)\right]+\right.\nonumber\\&&\left.
\frac{m^2_\pi-m^2_K}{m^2}\left[C_2(m^2_\pi,m^2,m^2_K)-C_2(m^2_\pi,0,m^2_K)\right]-C_2(s_1,m^2,m^2_K)-C_2(m^2_\pi,m^2,m^2_K)+
\right.\nonumber\\&&\left.C_1(s_2,m^2_K)+\ln m^2-1\right\}.
\end{eqnarray}
Here,
\begin{eqnarray}\label{C123}
C_1(s_2,m^2_K)&=&\int_0^1dx\ln[m^2_K-s_2x(1-x)-i\varepsilon],\nonumber\\
C_2(s_1,m^2,m^2_K)&=&\int_0^1dx\ln[(1-x)(m^2_K-s_1x)+m^2x-i\varepsilon],\\
C_3(s_1,s_2,m^2,m^2_K)&=&\int_0^1dx_1\int_o^{x_1}\frac{dx_2}{m^2_K+x_2(m^2-m^2_K)-(x_1-x_2)[s_2(1-x_1)+m^2_\pi
x_2]-s_1x_2(1-x_2)-i\varepsilon}.
\end{eqnarray}
\end{widetext}
We use the analytical expression for $C_1$ and $C_2$, while $C_3$
is evaluated numerically.  Note that in the kinematical region of
our interest the net contribution from the two-propagator terms
$C_{1,2}$ is negligible in comparison with the pure triangle
contribution $\propto C_3$, where all three poles are essential.
The knowledge of the explicit imaginary parts of the amplitude
(the discontinuities on the $K^*\bar K$, $\bar K^*K$, and  $K\bar
K$ cuts) permits one to control the result of numerical
evaluations. In the case of  one of the four charge modes they
look like
\begin{widetext}
\begin{eqnarray}\label{im1}
{\rm Im}g^{(K^{*+}K^-)}_{\iota f_0\pi}(m^2)=\frac{1}{2i}{\rm
Disc}_{K^{*+}K^-}(m^2)=\frac{g_{\iota
K^{*+}K^-}g_{K^{*+}K^+\pi^0}g_{f_0K^+
K^-}}{32\pi\sqrt{s_1}|p_\pi|}\left\{-4|p_\pi||p_
K|\left(1+\frac{s_1-m^2_{K^-}}{m^2}\right)\right. \nonumber\\ 
\left.+\left[s_1+m^2_\pi+2m^2_{K^-}-m^2-2s_2+\frac{(s_1-m^2_{K^-})
(m^2_{K^-}-m^2_\pi)} {m^2}\right]\,
\ln\frac{a_{K^{*+}K^-}+1+i\varepsilon}{a_{K^{*+}K^-}-1+i\varepsilon}\right\},
\qquad\end{eqnarray} where $a_{K^{*+}K^-}=(2E_{f_0}E_{
K^-}-s_2)/(2|p_\pi||p_{K^-}|)$,
$\,E_{f_0}=(s_1+s_2-m^2_\pi)/(2\sqrt{s_1})$,
$\,E_{K^-}=(s_1+m^2_{K^-}-m^2)/(2\sqrt{s_1})$,
$\,|p_\pi|=|p_{f_0}|=\sqrt{E^2_{f_0}-s_2}$, $\,|p_{\bar
K}|=\sqrt{E^2_{K^-}-m^2_{K^-}}$ (here, the mass $m$ of the $K^*$
meson is not fixed to be $m_{K^*}$);
\begin{eqnarray}\label{im2}
{\rm Im}g^{(K^+K^-)}_{\iota f_0\pi}(m^2)=\frac{1}{2i}{\rm
Disc}_{K^+K^-}(m^2)=\frac{g_{\iota
K^{*+}K^-}g_{K^{*+}K^+\pi^0}g_{f_0K^+
K^-}}{32\pi\sqrt{s_2}|p'_\pi|}\biggl\{4|p'_\pi||p'_
K|+\biggl[s_1+m^2_\pi+2m^2_{K^-}-m^2-2s_2\biggr.\biggr.\nonumber\\ 
\left.\left.+\frac{(s_1-m^2_{K^-})(m^2_{K^-}-m^2_\pi)}
{m^2}\right]\,
\ln\frac{a_{K^+K^-}+1-i\varepsilon}{a_{K^+K^-}-1-i\varepsilon}-
\frac{(s_1-m^2_{K^-})(m^2_{K^-}-m^2_\pi)}{m^2}\,
\ln\frac{a^{(0)}_{K^+K^-}+1}{a^{(0)}_{K^+K^-}-1}\right\},\ \
\end{eqnarray}
where  $a_{K^+K^-}\equiv a_{K^+K^-}(m^2)=-(2E'_\pi E'_{
K^-}+m^2_{K^-}+m^2_\pi-m^2)/(2|p'_\pi||p'_{K^-}|)$;
$\,E'_\pi=(s_1-s_2-m^2_\pi)/(2\sqrt{s_2})$,
$\,E'_{K^-}=\sqrt{s_2}/2$, $\,|p'_\pi|=\sqrt{E'^{\,2}_\pi-m^2_\pi}$,
$\,|p'_{K^-}|=\sqrt{E'^{\,2}_{K^-}-m^2_{K^-}}$;
$a^{(0)}_{K^+K^-}=a_{K^+K^-}(m^2=0)$.\end{widetext} The Lorenz
transformation from the $\iota$ rest frame to $f_0$ one gives the
relation $\sqrt{s_1}|p_\pi|=\sqrt{s_2}|p^\prime_\pi|$, so that the
coefficients in front of two logarithms originating from the
$K^\ast\bar K$, Eq.~(\ref{im1}), and $K\bar K$, Eq.~(\ref{im2}),
cuts are coincident. Hence, in the kinematical region where
imaginary parts of these logarithms appear, they cancel each other
due to different signs in front of $\varepsilon$. The logarithm with
$a^{(0)}_{K^+K_-}$ is explicitly real. So, the imaginary part of the
coupling constant
\begin{equation}\label{imtot}
{\rm Im}g_{\iota f_0\pi}(m^2)=\frac{1}{2i}\left[{\rm Disc}_{K^{*+}
K^-}(m^2)+{\rm Disc}_{K^+K^-}(m^2) \right]\end{equation} is real.
We have verified that the imaginary part of the numerically
evaluated triangle diagram coincides with the evaluation of the
analytically calculated one.

To account for the effect of the finite $K^*$ width, we write the
propagator of the unstable $K^*$ meson in the form of the spectral
K\"{a}ll\'{e}n-Lehmann representation \cite{AK1,AK2,AK3}
\begin{equation}\label{KLrep}
\frac{1}{m^2_{K^*}-p_{K^*}^2-im_{K^*}\Gamma_{K^*}}\to\int\limits^{\infty}_{
(m_{K}+m_\pi)^2}dm^2\frac{\rho(m^2)}{m^2-p_{K^*}^2-i\varepsilon}
\end{equation} and approximate $\rho(m^2)$ in the following way:
\begin{equation}\label{Eq1-16}
\rho(m^2)=\frac{1}{\pi}\frac{m_{K^*}\Gamma_{K^*}}
{(m^2-m^2_{K^*})^2+(m_{K^*}\Gamma_{K^*})^2}\,.
\end{equation}
Then, instead of amplitude $T\equiv T(s_1,s_2,m^2,m^2_K)$ from Eq.
\ref{tri1},  we have the amplitude $\langle T\rangle$ weighted with
the spectral density $\rho(m^2)$ \cite{AK1,AK2,AK3}
\begin{equation}\label{WT}
\langle T\rangle=\int\limits^{\infty}_{
(m_{K}+m_\pi)^2}\rho(m^2)\,T(s_1,s_2,m^2,m^2_K)\,dm^2.
\end{equation}
This integration eliminates the logarithmic infinities in the
imaginary part of the triangle diagram. Notice that the
contributions of the discontinuities on the $K^*\bar K$ and $\bar
K^*K$ cuts in the $s_1$ channel are caused by the real three-body
intermediate states $K\pi\bar K$ and $\bar K\pi K$, respectively.
At the same time, the discontinuities of the triangle diagram in
the $s_2$ channel correspond to the two-body intermediate states
$K\bar K$.

The amplitude of the subprocess $K^+ K^-\to f_0(980)\to\pi^+\pi^-$
(or $K^0\bar K^0\to f_0(980)\to\pi^+ \pi^-$), being a part of the
amplitude of the diagram in Fig.~\ref{FigEta3pi}, is taken in the
form
\begin{equation}\label{Eq5}
f_S(s_2)=\frac{g_{f_0K^+K^-} g_{f_0 \pi^+\pi^-}}{16\pi}\,\frac{1}
{D_{f_0}(s_2)}\, e^{i\varphi(s_2)}\,,
\end{equation} where $g_{f_0K^+K^-}$ $(=g_{f_0K^0\bar K^0})$ and  $g_{
f_0\pi^+\pi^-}$ $(=\sqrt{2}g_{ f_0\pi^0\pi^0})$ are the coupling
constants of $f_0(980)$ with  $K^+K^-$ ($K^0\bar K^0$) and
$\pi^+\pi^-$ ($\pi^0\pi^0$), the phase of the background is
$\varphi(s_2)\approx \pi/2$, and $1/D_{f_0}(s_2)$ stands for the
$f_0(980)$ propagator \cite{ADS80}, the expression of which takes
into account the couplings of $f_0(980)$ with the  $\pi\pi$ and
$K\bar K $ channels and the corresponding finite width
corrections,
\begin{equation}\label{Eq5-a}
\frac{1}{D_{f_0}(s_2)}=\frac{1}{m^2_{f_0}
-s_2+\sum_{ab}[\mbox{Re}\Pi^{ab}_{f_0}(m^2_{f_0})-\Pi^{ab}_{f_0}(s_2)
]}.\end{equation} Here, $\Pi^{ab}_{f_0}(s_2)$ is the polarization
operator for the $f_0(980)$, corresponding to the contribution of
the $ab$ intermediate state ($ab=\pi^+\pi^-,\,\pi^0\pi^0,\,
K^+K^-,\,K^0\bar K^0$); $\mbox{Im}\Pi^{ab}_{f_0}(s_2)/\sqrt{s_2}
=\Gamma_{f_0\to ab}(s_2)=g^2_{f_0ab}\rho_{ab}(s_2)/(16\pi)$ is the
width of the $f_0(980)\to ab$ decay; in this case $m_a=m_b$, and
for $s_2>4m^2_a$
\begin{equation}\label{Eq5-b} \Pi^{ab}_{f_0}(s)=\frac{g^2_{f_0
ab}}{16\pi}\rho_{ab}(s)\left[i-\frac{1}{\pi}\ln\frac{1+
\rho_{ab}(s_2)}{1-\rho_{ab}(s_2)}\right]\,,\label{Pi-f01}\end{equation}
where $\rho_{ab}(s_2)=\sqrt{1-4m_a/s_2}$; for
0\,$<$\,$s_2$\,$<$\,$4m^2_a$, $\rho_{ab}(s_2)$ should be replaced by
$i|\rho_{ab}(s_2)|$ and
\begin{equation}\label{Eq5-c}\Pi^{ab}_{f_0}(s_2)=-\frac{g^2_{f_0
ab}}{16\pi}|\rho_{ab}(s_2)|\left[1-\frac{2}{\pi}\arctan
|\rho_{ab}(s_2)|\right].\label{Pi-f02}\end{equation}

Our estimates are given for the following values: $m_{f_0}=0.990$
GeV, $2g^2_{f_0K^+K^-}/(16\pi)=0.4$ GeV$^2$, and
$(3/2)g^2_{f_0\pi^+\pi^-}/(16\pi)=0.1$ GeV$^2$. We have also tried
different values of the $f_0(980)$ parameters, for instance,
$m_{f_0}=0.975$ GeV, $2g^2_{f_0K^+K^-}/(16\pi)=0.5$ GeV$^2$, and
$(3/2)g^2_{f_0 \pi^+\pi^-}/(16\pi)=0.1$ GeV$^2$ and verified that
the results are not changed significantly.

For the example given in Fig.~\ref{SpecEtapipi}, the following
values are used for the $a_0(980)$ resonance \cite{ADS80,ADS80a}:
$m_{a_0}=0.9847$ GeV, $2g^2_{a_0K^+K^-}/(16\pi)=0.4$ GeV$^2$, and
$g^2_{a_0\eta\pi}$\,=$\,g^2_{a_0\eta'\pi}$\,=$\,g^2_{a_0K^+K^-}$.
To take into account the $\pi\pi$ final state interaction in the
decay $\iota\to\eta\pi\pi$, the contribution of the amplitude
$\iota\to a_0(908)\pi\to\eta(\pi\pi)_S$ is multiplied by the
factor $[1+i\rho_{\pi\pi}(s)T^0_0(s)]=
e^{i\delta^0_0(s)}\cos\delta^0_0 (s)$, where $(\pi\pi)_S$ means
the $\pi\pi$ system in $S$ wave, $T^0_0(s)$ and $\delta^0_0(s)$
being, respectively,  the amplitude and the phase of $\pi\pi$
scattering with the angular momentum $l=0$ and isospin $I=0$, $s$
is the invariant mass squared of the $\pi\pi$ state. The data on
$\delta^0_0(s)$ are approximated by the smooth curve
\cite{AS07,AS94}.

The propagator of the $a_0(980)$ resonance with the invariant mass
square $s_2$ is
\begin{equation}\label{Eq5-d}
\frac{1}{D_{a_0}(s_2)}=\frac{1}{m^2_{a_0}
-s_2+\sum_{ab}[\mbox{Re}\Pi^{ab}_{a_0}(m^2_{f_0})-\Pi^{ab}_{a_0}(s_2)
]},\end{equation} where $ab=\pi\eta,\,K^+K^-,\,K^0\bar K^0,\,\pi
\eta'$; $\mbox{Im}\Pi^{ab}_{a_0}(s_2)/\sqrt{s_2} =\Gamma_{a_0\to
ab}(s_2)=g^2_{a_0ab}\rho_{ab}(s_2)/(16\pi)$. For
$s_2>m_{ab}^{(+)\,2}$ ($m_{ab}^{(\pm)}$\,=\,$m_b\pm m_a$, $m_b\geq
m_a$), the polarization operator is given by \cite{ADS80,ADS80a}
\begin{eqnarray}\label{Pi-ab-a01}\Pi^{ab}_{a_0}(s_2)=\frac{g^2_{a_0
ab}} {16\pi}\left[\frac{m_{ab}^{(+)}m_{ab}^{(-)}}{\pi
s_2}\ln\frac{m_a}{m_b}+\rho_{ab}(s_2)\right.\ \nonumber\\
\left.\times\left(i-\frac{1}{\pi}\,\ln\frac{
\sqrt{s_2-m_{ab}^{(-)\,2}}+\sqrt{s_2-m_{ab}^{(+)\,2}}}
{\sqrt{s_2-m_{ab}^{(-)\,2}}-\sqrt{s_2-m_{ab}^{(+)\,2}}}
\right)\right],\end{eqnarray} where
$\rho_{ab}(s_2)$\,=\,$\sqrt{s_2-m_{ab}^{(+)\,2}}\,\sqrt{
s_2-m_{ab}^{(-)\,2}}\Big/s_2$, for
$m_{ab}^{(-)\,2}<s_2<m_{ab}^{(+)\,2}$
\begin{eqnarray}\label{Pi-ab-a02}\Pi^{ab}_{a_0}(s_2)=\frac{g^2_{a_0ab}}
{16\pi}\left[\frac{m_{ab}^{(+)}m_{ab}^{(-)}}{\pi
s_2}\ln\frac{m_a}{m_b}\right.\nonumber\\
\left.-\rho_{ab}(s_2)\left(1-\frac{2}{\pi}\arctan\frac{\sqrt{
m_{ab}^{(+)\,2}-s_2}}{\sqrt{s_2-m_{ab}^{(-)\,2}}}\right)\right],\end{eqnarray}
where $\rho_{ab}(s_2)$\,=\,$\sqrt{m_{ab}^{(+)\,2}-s_2}
\,\sqrt{s_2-m_{ab}^{(-)\,2}}\Big/s_2$, and for $s_2\leq
m_{ab}^{(-)\,2}$
\begin{eqnarray}\label{Pi-ab-a03}\Pi^{ab}_{a_0}(s_2)=\frac{g^2_{a_0
ab}}{16\pi}\left[\frac{m_{ab}^{(+)}m_{ab}^{(-)}}{\pi
s_2}\ln\frac{m_a}{m_b}\right.\ \ \nonumber\\
\left.-\rho_{ab}(s_2)\frac{1}{\pi}\,\ln\frac{
\sqrt{m_{ab}^{(+)\,2}-s_2}+\sqrt{m_{ab}^{(-)\,2}-s_2}}
{\sqrt{m_{ab}^{(+)\,2}-s_2}-\sqrt{m_{ab}^{(-)\,2}-s_2}}\right],\end{eqnarray}
where $\rho_{ab}(s_2)$\,=\,$\sqrt{m_{ab}^{(+)\,2}-s_2}\,\sqrt{
m_{ab}^{(-)\,2}-s_2}\Big/s_2$.

\section{Conclusion}

The phenomenon of the $a^0_0(980)-f_0(980)$ mixing \cite{ADS79}
gave an impetus to conduct experiments of VES on the decay
$f_1(1285)$\,$\to$\,$ \pi^+\pi^-\pi^0$ \cite{Do08,Do11} and BESIII
on the decays $J/\psi \to\phi f_0(980)\to\phi
a_0(980)\to\phi\eta\pi$, $\,\chi_{c1} \to a_0(980)\pi^0\to
f_0(980)\pi^0\to\pi^+\pi^-\pi^0$ \cite{Ab1}, and
$J/\psi\to\gamma\eta(1405)\to\gamma f_0(980)\pi^0\to\gamma3\pi$
\cite{Ab2}. We hope that the remarks presented here, on the
mechanisms of the isospin breaking in the decay
$\eta(1405)\to3\pi$, will stimulate both the further studies of
this decay and the principal improvement of the data about $K\bar
K$, $ K\pi$, $\eta\pi$, and $\pi\pi$ mass spectra in the decays of
the resonance structure $\eta(1405/1475)$ into $K\bar K\pi$ and
$\eta\pi\pi$, and about the shape of these resonance peaks in the
$K\bar K\pi$ and $\eta\pi\pi$ channels. \vspace{0.1cm}

The present work is partially supported by the Russian Foundation
for Basic Research Grant No. 13-02-00039 and by the
interdisciplinary project Grant No. 102 of the Siberian Branch of
Russian Academy of Sciences.



\begin{thebibliography}{99}
\bibitem{FN2} In the region between the $K^+K^-$ and $K^0\bar K^0$
thresholds the isospin breaking decay amplitude has the following
effective structure:
$$\left|\ln\left(\frac{\Gamma_{K^*}/2}
{\sqrt{m^2_{K^0}-m^2_{K^+}+\Gamma^2_{K^*}/4}}\right)\right|.$$
This expression is not a result of calculations. It is
constructed, so to speak, by hand to reflect the properties of
this amplitude.
\bibitem{FN3}
As for the logarithmic singularity, higher-order corrections in
the effective Lagrangian approach are reduced to renormalizations
of the coupling constants and so are included into their physical
values. The sum of the triangle diagrams with charged and neutral
$K\bar K^*+\bar KK^*$ intermediate states gives the primary
isospin violation contribution. Other isospin violation
corrections are not essential since they are proportional to
$m_{K^0}-m_{K^+}$.
\end{thebibliography}
\end{document}